\documentclass{svproc}
\usepackage{url}

\usepackage{graphicx}   
\usepackage{float}    
\usepackage{multicol}        
\usepackage[bottom]{footmisc}

\usepackage{lineno}

\usepackage{blindtext}

\begin{document}
\mainmatter              
\title{Quarkonium measurements at forward rapidity with ALICE at the LHC}
\titlerunning{Quarkonium at forward rapidity with ALICE }  
%
\author{Wadut Shaikh (for the ALICE Collaboration)}
\institute{ Saha Institute of Nuclear Physics, Kolkata-700064, India\\
\email{wadut.shaikh@cern.ch}
}

\maketitle              
 \vspace{-0.2cm}
\begin{abstract}
Heavy quarks are produced at the first instant of a nucleus­--nucleus collision and therefore are an important tool to study the subsequent high energy­-density medium formed in ultra-­relativistic heavy­-ion collisions. A series of experimental efforts for understanding the properties of the Quark­--Gluon Plasma (QGP), a medium consisting of a deconfined state of quarks and gluons, are based on measuring the bound states of heavy quark­-antiquark pairs known as quarkonia. However, the medium modification of heavy­-flavour hadron production includes also the contribution of Cold Nuclear Matter (CNM) effects such as shadowing or nuclear break­up in addition to the QGP effects. Proton­--nucleus collisions, where no QGP is expected, are used to measure CNM effects on quarkonium production. Finally, quarkonium measurements in proton­--proton collisions are used as reference for both heavy­-ion and proton­--ion collisions. ALICE measurements of quarkonia at forward rapidity for various energies and colliding systems (pp, p­--Pb, Pb­--Pb and Xe­--Xe) during the LHC Run­ 1 and Run­ 2 periods will be discussed. 
\end{abstract}
 \vspace{-0.7cm}
\section{Introduction}

Quarkonia are a fruitful probe to investigate the properties of the deconfined medium, called Quark--Gluon Plasma (QGP), created in ultra-relativistic heavy-ion collisions. At LHC energies the modification of the J/$\psi$ ($c$$\bar{c}$) in heavy-ion collisions with respect to the binary-scaled yield in pp collisions is explained as an interplay of suppression~\cite{Matsui} and (re)generation~\cite{regenaration1,regenaration2}. For $\Upsilon$ ($b$$\bar{b}$) on the other hand, (re)generation effects are expected to be negligible due to the small number of produced $b$ quarks. In addition, the regenerated quarkonia inherit the flow of their constituting $c$ quarks and thus participate to the collective motion in the QGP. The CNM effects (shadowing, parton energy loss, interaction with hadronic medium) which are not related to the deconfined medium may also lead to a modification of quarkonium production. In order to disentangle the CNM effects from the hot nuclear matter effects, quarkonium production is studied in p--Pb collisions in which the QGP is not expected to be formed. In pp collisions, the quarkonium production can be understood as the creation of a heavy-quark pair ($q\bar{q}$) (perturbative) followed by its hadronization into a bound state (non-perturbative). None of the existing models fully describe the quarkonium production in pp collisions and more differential measurements will further constrain the quarkonium production models in elementary hadronic collisions.\\
 \vspace{-0.9cm}
\section{Analysis and results}
The ALICE Collaboration has studied quarkonium production in various collision systems (pp, p--Pb, Pb--Pb, and Xe--Xe) at different center-of-mass energies per nucleon pair $\sqrt{s_{\rm NN}}$ down to zero transverse momentum ($p_{\rm T}$) and at forward rapidity ($2.5<y<4$) with the Muon Spectrometer~\cite{MCH} through the dimuon decay channel.

\begin{figure}[htb]
\begin{center}
\includegraphics[height=5.0cm,,angle=0]{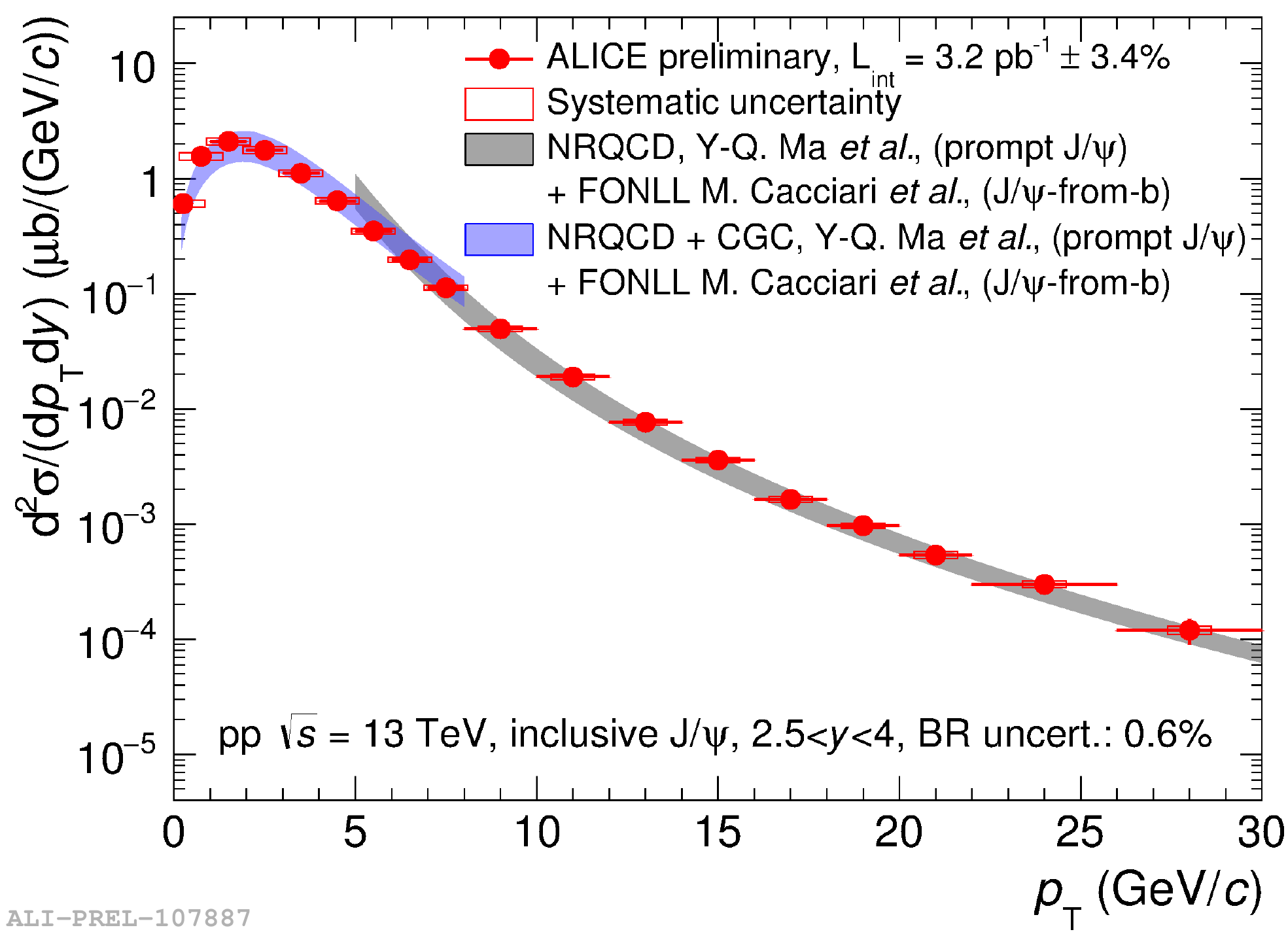}~~~
\includegraphics[height=5.0cm,,angle=0]{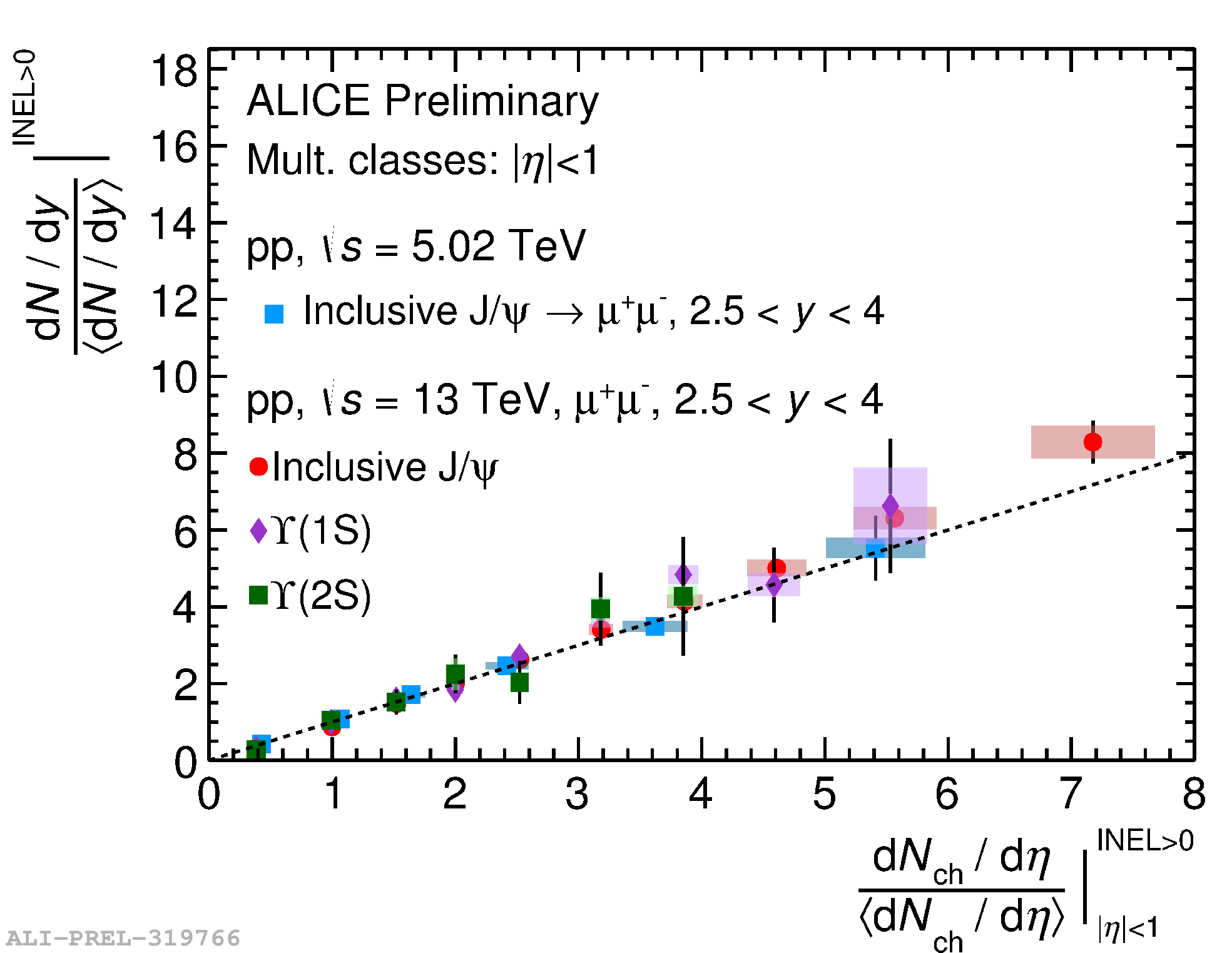}
\caption{$p_{\rm T}$ differential inclusive J/$\psi$ cross section measured at forward rapidity in pp collisions at $\sqrt{s}$ = 13 TeV (left). Relative quarkonium yield as a function of the relative charged-particle density in pp collisions at $\sqrt{s}$ = 13 TeV and 5.02 TeV (right).  }
\label{fig1}
\end{center}
\end{figure}

 \hspace{-0.6cm} {\bf pp collisions}: 
  In Fig.~\ref{fig1} (left), the inclusive J/$\psi$ production cross section at $\sqrt{s}$ = 13 TeV is
compared to different sets of Non-Relativistic QCD
(NRQCD) predictions~\cite{pp13TeV}. Recent theoretical developments, e.g. combining NRQCD with Color Glass Condensate (GCG+NRQCD) ~\cite{pp13TeVModel} reproduce the J/$\psi$ $p_{\rm T}$ shape. In Fig.~\ref{fig1} (right), the quarkonium production shows a linear increase with charged-particle multiplicity. 
 
 \vspace{-0.6cm}

\vspace{-1.0cm}

The ALICE Collaboration has also measured the inclusive J/$\psi$ polarization in pp collisions at $\sqrt{s}$ = 7 TeV and 8 TeV~\cite{ppPolarisation}. The measured J/$\psi$ polarization is compatible with zero within uncertainties in the different studied $p_{\rm T}$ intervals. The Color-Singlet Model (CSM) and the Next-to-Leading Order (NLO) NRQCD~\cite{NLOpolaJpsi} predictions do not describe the polarization parameters. \\

\hspace{-0.6cm} {\bf p--Pb collisions}: The CNM effects can be studied in p--Pb collisions via the nuclear modification factor ($R_{\rm pA}$) defined as \\
\vspace{-0.1cm}
$$ R_{\rm pPb} = \frac{\sigma_{\rm pPb}}{A_{\rm Pb}~.~\sigma_{\rm pp}},$$
\vspace{-0.2cm}

where $\sigma_{\rm pPb}$ and $\sigma_{\rm pp}$ are the production cross sections in p--Pb and pp collisions respectively. $A_{\rm Pb}$ is the atomic mass number (208) of the
Pb nucleus. Inclusive J/$\psi$ production in p--Pb collisions at $\sqrt{s_{\rm NN}}$ =  8.16 TeV~\cite{pPbJpsi8} was measured by ALICE and a suppression was observed at positive $y_{cms}$ corresponding to low bjorken-x gluons in the Pb nucleus. The models which include various combinations of cold nuclear matter effects describe the J/$\psi$ p--Pb data.

\vspace{-0.6cm}
\begin{figure}[htb]
\begin{center}
\includegraphics[height=5.0cm,,angle=0]{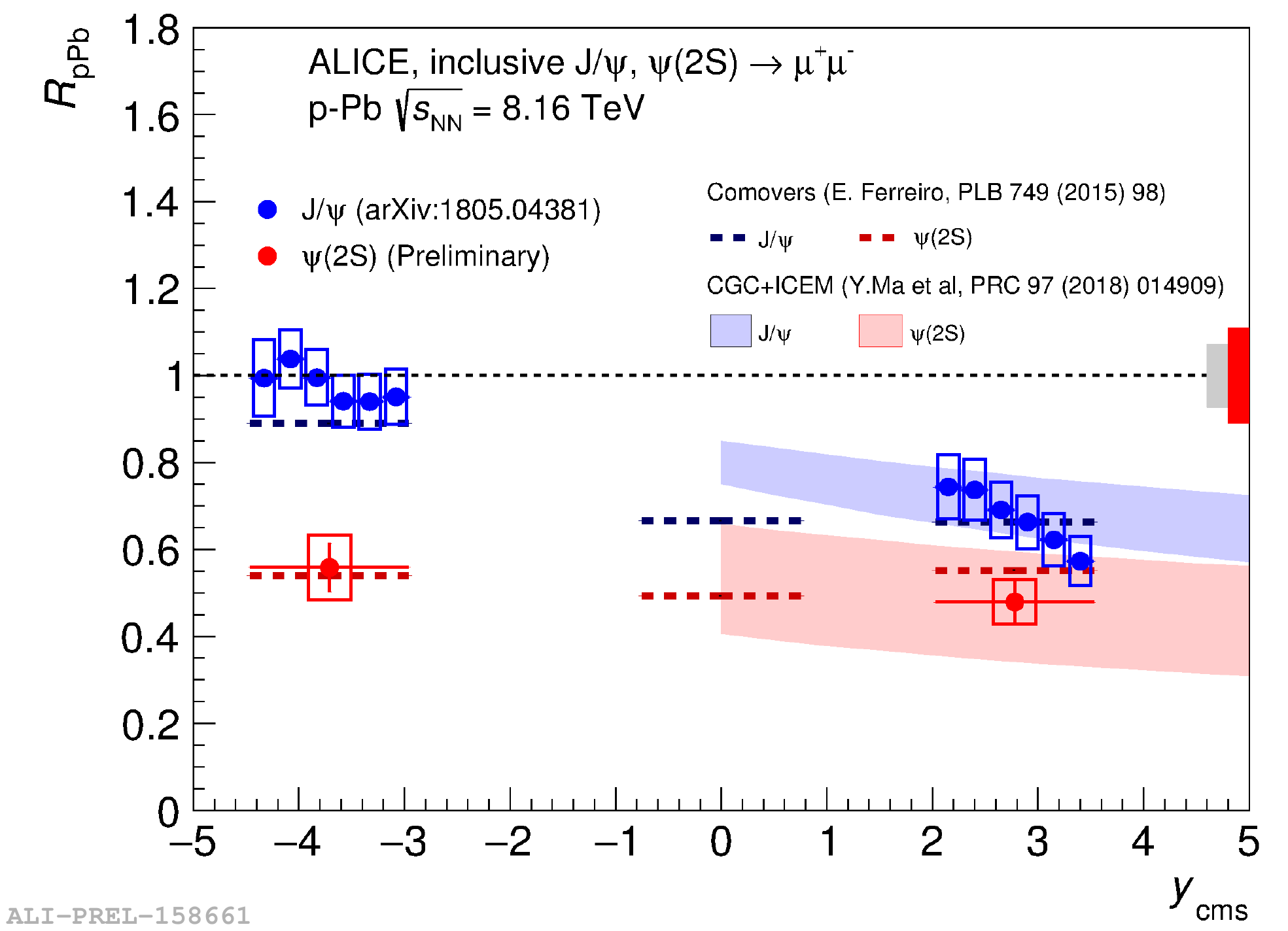}~~~
\includegraphics[height=5.0cm,,angle=0]{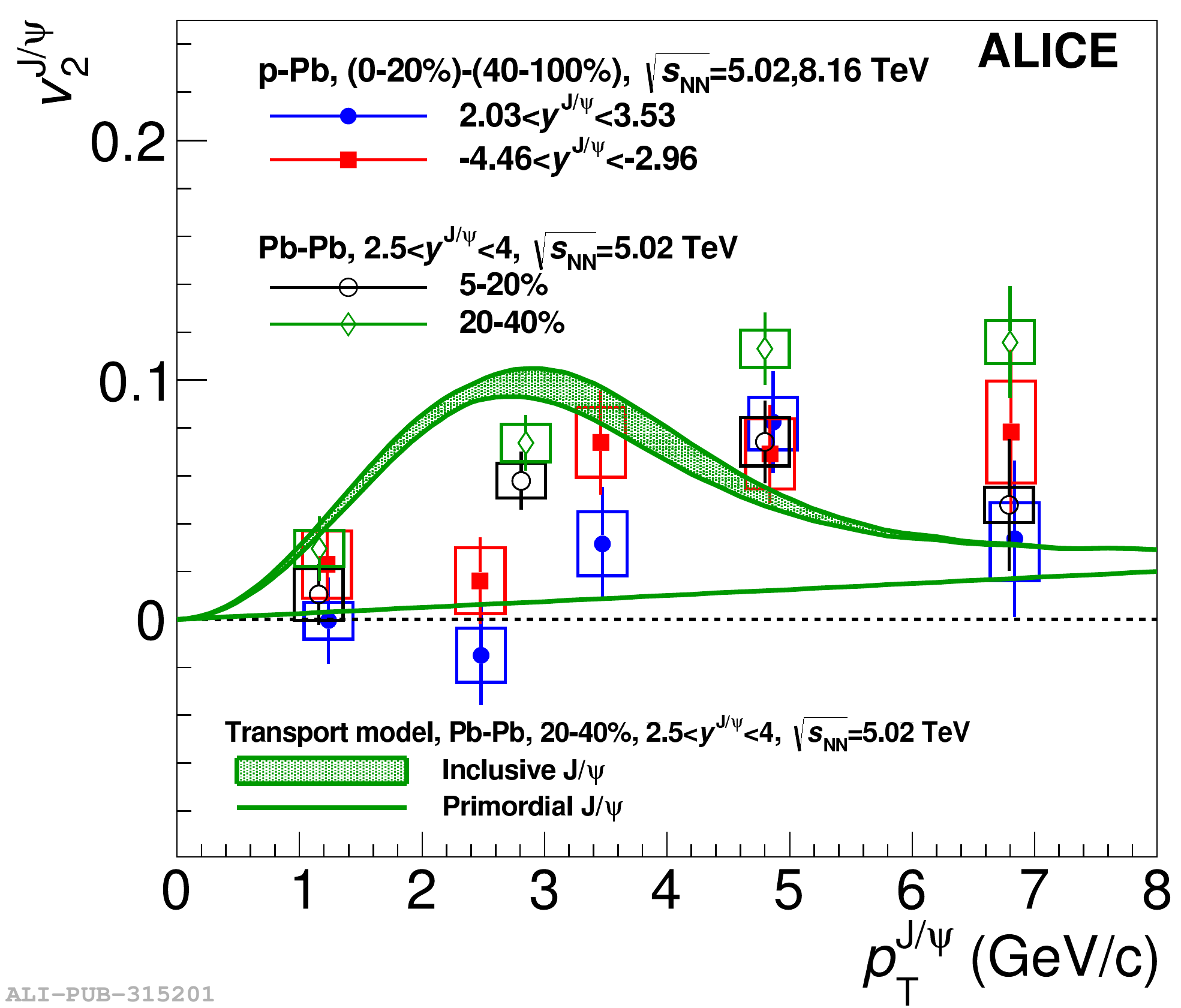}
\caption{ The $\psi$(2S) and J/$\psi$ $R_{\rm pA} $ as a function of $y_{\rm cms}$ with different model predictions in p--Pb collisions at $\sqrt{s_{\rm NN}}$ =  8.16 TeV (left). J/$\psi$ $v_{2}$ coefficients in p--Pb and Pb--p collisions compared to the results in Pb--Pb at $\sqrt{s_{\rm NN}}$ = 5.02 TeV and the transport model calculations for semi-central Pb--Pb collisions at $\sqrt{s_{\rm NN}}$ = 5.02 TeV (right).  }
\label{fig2}
\end{center}
\end{figure}

\vspace{-1cm}

The $R_{\rm pA}$ of $\psi$(2S) together with J/$\psi$ is shown in Fig.~\ref{fig2} (left). Contrary to the J/$\psi$ case~\cite{pPbJpsi8}, models must include final state interactions with the surrounding medium in order to describe the $\psi$(2S) results.\\

\vspace{-0.25cm}
The ALICE Collaboration measured azimuthal correlations between J/$\psi$ emitted at forward and backward rapidities with mid-rapidity charged particles at $\sqrt{s_{\rm NN}}$ = 5.02 and 8.16 TeV ~\cite{pPbJpsiV2}. The data indicate persisting long-range correlation structures at $\Delta \varphi$ $\approx$ 0 and $\Delta \varphi$ $\approx$ $\pi$. The elliptic flow is defined as the $2^{nd}$ order
coefficient $v_{2}$ of the Fourier expansion of the
azimuthal distribution. Fig.~\ref{fig2} (right) shows the elliptic flow $v_{2}$ of J/$\psi$ for p--Pb and Pb--p collisions~\cite{pPbJpsiV2} together with measurements~\cite{PbPbJpsiv2} and model calculations for Pb--Pb collisions at $\sqrt{s_{\rm NN}}$ = 5.02.  The J/$\psi$ $v_{2}$ in $3<p_{\rm T}<6$ GeV/$c$ is found to be positive in both rapidity intervals and of the same order as that measured in Pb--Pb collisions.\\

\begin{figure}[htb]
\begin{center}
\includegraphics[height=5.0cm,,angle=0]{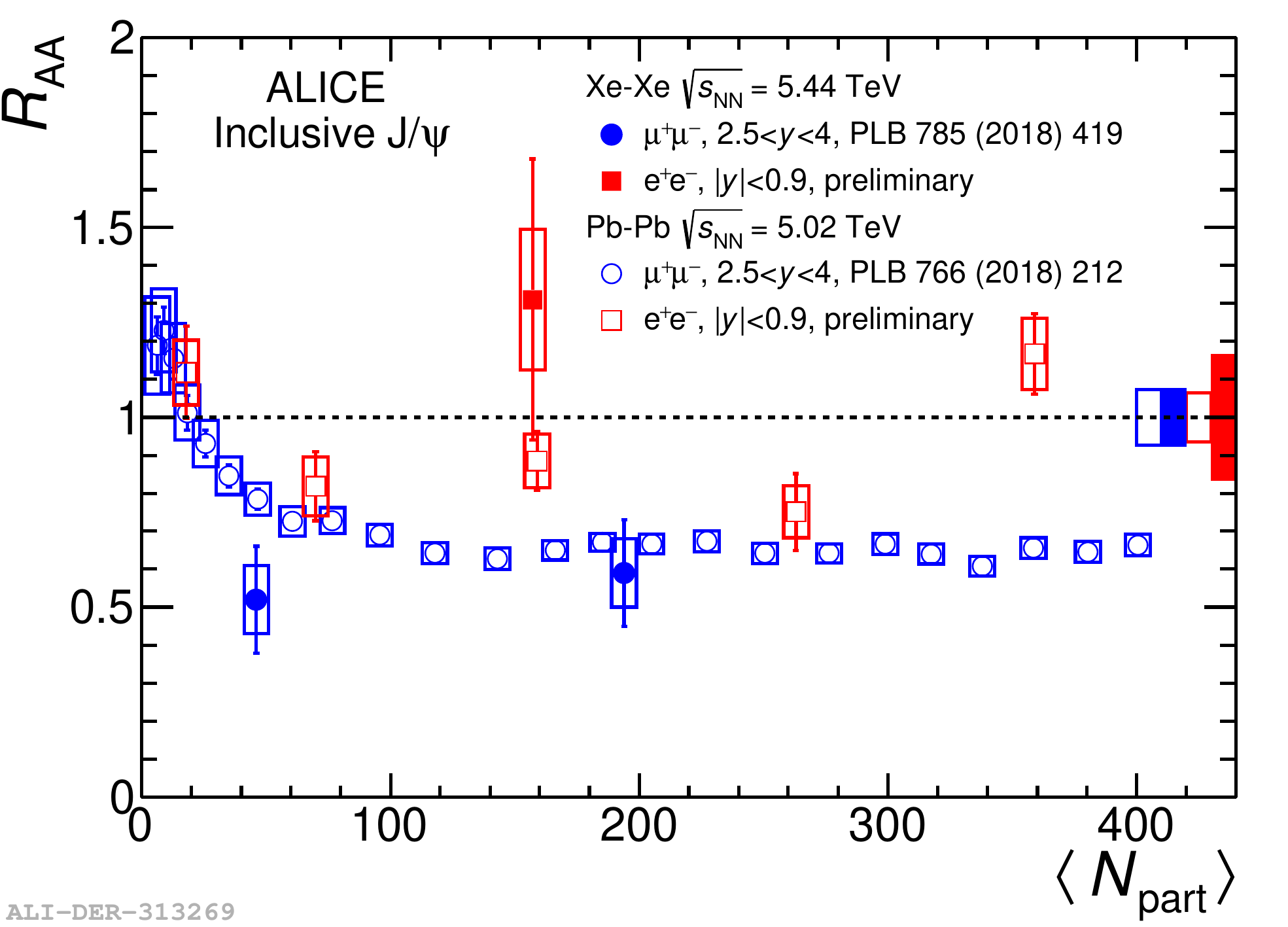}~~~
\includegraphics[height=5.0cm,,angle=0]{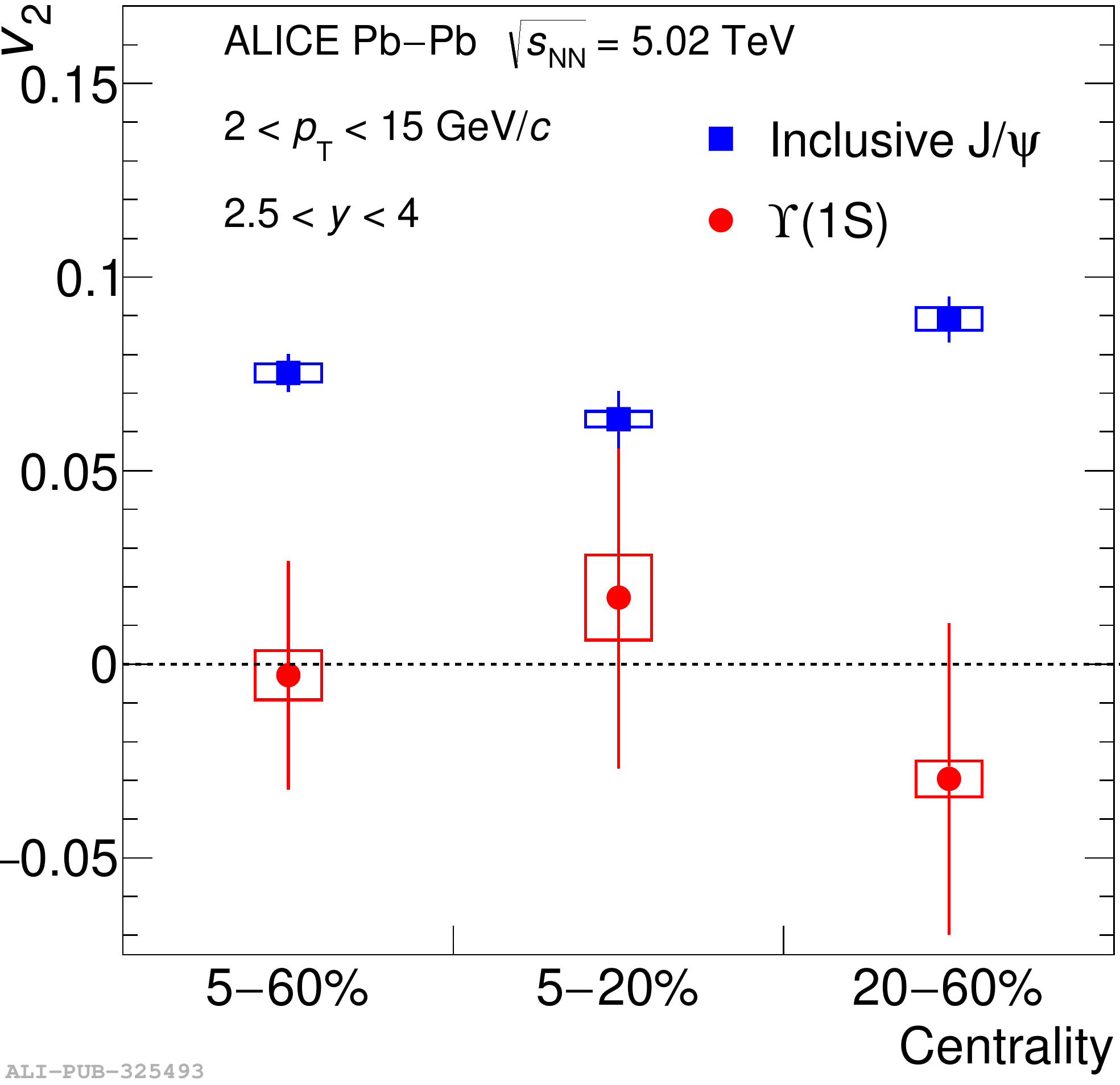}
\caption{ J/$\psi$ $R_{\rm AA}$ in Pb--Pb (Xe--Xe) at $\sqrt{s_{\rm NN}}$ = 5.02 (5.44) TeV (left). The $\Upsilon$(1S) $v_{2}$ coefficient integrated over the transverse momentum range $2<p_{\rm T}<15$ GeV/$c$ in three centrality intervals compared to that of inclusive J/$\psi$ at $\sqrt{s_{\rm NN}}$ = 5.02 TeV (right). }
\label{fig3}
\end{center}
\end{figure}

\hspace{-0.4cm} {\bf Pb--Pb (Xe--Xe) collisions}: The nuclear modification
factor for a given centrality
class $i$ in A--A collisions can be defined as
\vspace{-0.2cm}
$$  R^{i}_{\rm AA} = \frac{{\rm d}^2N^{{\rm AA}}_{i}/{\rm d}y{\rm d}p_{\rm T}}{<T^i_{\rm {AA}}>~.~{\rm d}^2\sigma^{\rm{pp}}/{\rm d}y{\rm d}p_{\rm T}}, $$
\vspace{-0.2cm}


where ${\rm d}^2N^{{\rm AA}}_{i}/{\rm d}y{\rm d}p_{\rm T}$ is the yield in nucleus-nucleus collisions, $<T^i_{\rm{AA}}>$ is the nuclear overlap function and ${\rm d}^2\sigma^{\rm{pp}}/{\rm d}y{\rm d}p_{\rm T}$ is the production cross section in pp collisions. In Fig.~\ref{fig3} (left), the J/$\psi$ $R_{\rm AA}$ measured in Xe--Xe collisions~\cite{RAA_Jpsi_XeXe} as a function of centrality is compared to the one measure in Pb--Pb collisions~\cite{RAA_Jpsi_PbPb} both at forward rapidity and mid-rapidity ($-0.5<y<0.5$).  The ALICE collaboration also measured the $R_{\rm PbPb}$ of $\Upsilon$ states  for Pb--Pb collisions at $\sqrt{s_{\rm NN}}$ = 5.02~\cite{PbPbUpsilon}. A strong suppression in central collisions is observed for $\Upsilon$(1S). A larger suppression of $\Upsilon$(2S) compared to $\Upsilon$(1S) is also observed.

  \vspace{-0.6cm} 

 \vspace{-0.9cm}

The data samples recorded by ALICE during the 2015 and 2018 LHC Pb--Pb runs at $\sqrt{s_{\rm NN}}$ = 5.02 TeV are used for $\Upsilon$(1S) $v_{2}$ 
measurement~\cite{UpsilonV2}. In Fig.~\ref{fig3} (right), the $v_{2}$ coefficient of $\Upsilon$(1S) in three centrality intervals is shown, together with that of the J/$\psi$~\cite{PbPbJpsiv2}. The measured $\Upsilon$(1S) $v_{2}$ coefficient is compatible with zero within current uncertainties and consistent with no recombination expectations in the bottomonium sector. The $R_{\rm AA}$ and $v_{2}$ results suggest that (re)generation is the dominant contribution to the production of J/$\psi$, but contributes only marginally or zero to the production of $\Upsilon$(1S) in heavy-ion collisions at LHC collision energies.

\end{document}